\documentclass[aps,prx,twocolumn,superscriptaddress]{revtex4}
\usepackage{amssymb}
\usepackage{amsfonts}
\usepackage{graphicx}
\usepackage{amsmath}
\usepackage{array}
\usepackage{lineno,hyperref}
\usepackage{footnote}
\begin{document}


\title{From weak to strong-coupling superconductivity tuned by substrate in TiN films}

\author{Yixin Liu}
\affiliation{State Key Laboratory of Materials for Integrated
Circuits, Shanghai Institute of Microsystem and Information
Technology, Chinese Academy of Sciences, Shanghai 200050,
China}\affiliation{University of Chinese Academy of Sciences, Beijing
100049, China}

\author{Yuchuan Liu}
\affiliation{State Key Laboratory of Materials for Integrated
Circuits, Shanghai Institute of Microsystem and Information
Technology, Chinese Academy of Sciences, Shanghai 200050,
China}\affiliation{Shanghai University,
Shanghai 200444, China}

\author{Zulei Xu}
\affiliation{State Key Laboratory of Materials for Integrated
Circuits, Shanghai Institute of Microsystem and Information
Technology, Chinese Academy of Sciences, Shanghai 200050,
China}\affiliation{University of Chinese Academy of Sciences, Beijing
100049, China}

\author{Aobo Yu}
\affiliation{State Key Laboratory of Materials for Integrated
Circuits, Shanghai Institute of Microsystem and Information
Technology, Chinese Academy of Sciences, Shanghai 200050,
China}\affiliation{University of Chinese Academy of Sciences, Beijing
100049, China}

\author{Xiaoni Wang}
\affiliation{State Key Laboratory of Materials for Integrated
Circuits, Shanghai Institute of Microsystem and Information
Technology, Chinese Academy of Sciences, Shanghai 200050,
China}\affiliation{University of Chinese Academy of Sciences, Beijing
100049, China}

\author{Wei Peng}
\affiliation{State Key Laboratory of Materials for Integrated
Circuits, Shanghai Institute of Microsystem and Information
Technology, Chinese Academy of Sciences, Shanghai 200050,
China}\affiliation{University of Chinese Academy of Sciences, Beijing
100049, China}

\author{Yu Wu}\email[]{wuyu@mail.sim.ac.cn}
\affiliation{State Key Laboratory of Materials for Integrated
Circuits, Shanghai Institute of Microsystem and Information
Technology, Chinese Academy of Sciences, Shanghai 200050, China}

\author{Gang Mu}
\email[]{mugang@mail.sim.ac.cn} \affiliation{State Key Laboratory
of Materials for Integrated Circuits, Shanghai Institute of
Microsystem and Information Technology, Chinese Academy of Sciences,
Shanghai 200050, China}\affiliation{University of Chinese Academy of
Sciences, Beijing 100049, China}

\author{Zhi-Rong Lin}
\email[]{zrlin@mail.sim.ac.cn} \affiliation{State Key Laboratory
of Materials for Integrated Circuits, Shanghai Institute of
Microsystem and Information Technology, Chinese Academy of Sciences,
Shanghai 200050, China}\affiliation{University of Chinese Academy of
Sciences, Beijing 100049, China}

\begin{abstract}
The interplay between substrates and superconducting thin films has attracted increasing attention. Here, we report an in-depth investigation on superconducting properties
of the epitaxial TiN thin films grown on three different substrates by dc reactive magnetron sputtering. The TiN films grown on (0001) sapphire exhibit (111) crystal orientation,
while that grown on (100) Si and MgO substrates exhibit (100)
orientation. Moreover, the samples grown on Si reveal a relatively lower level of disorder, accompanied by the higher critical transition temperature $T_c$ and smaller magnitude of upper critical field slope near $T_c$.
Remarkably, we uncovered a rather high value of superconducting gap (with $\Delta_0/k_BT_c$ = 3.05) in TiN film on Si indicating a very strong coupling superconductivity, in sharp contrast to the case using sapphires and MgO as the
substrate which reveals a weak-coupling feature. The comprehensive analysis considering the scenarios of the three substrate shows that
the grain size of the thin films may be an important factor influencing the superconductivity.\\

\end{abstract}

 \maketitle

\section*{1. Introduction}
Nitride superconducting thin films have excellent application performance in superconducting (SC) electronic device~\cite{YouLX2017,SQUID2017,QPS2022,Wang_2024}.
Due to the low
microwave loss~\cite{10.1063/1.3517252} and high kinetic
inductance~\cite{10.1063/1.4771995}, TiN films attracted more and more interests for the application in superconducting microwave
resonators~\cite{10.1063/1.4813269,6942188,Torgovkin_2018,10.1063/1.5053461} in recent years.
It is reported that quality factor higher than $2\times 10^6$ can be achieved in
microwave resonator made with TiN
films~\cite{10.1063/1.5053461,PhysRevMaterials.6.036202}.
Typically TiN films can be grown by magnetron
sputtering~\cite{10.1116/1.573255}, pulsed laser deposition (PLD)
~\cite{10.1063/1.107568,VISPUTE1998431}, and molecular beam epitaxy~\cite{10.1063/1.4759019}. It is found that physical performance of TiN
films can be tuned by the growth conditions like the chamber pressure and gas-flow rate~\cite{Torgovkin_2018}. The $T_c$ value of
TiN films grown on (100) MgO could be over 5 K, which is very close to
bulk material~\cite{10.1063/1.4759019}.

The physical properties of TiN thin films have also been widely studied.
The magnetic field induced superconductor-insulator transition has been
observed in TiN thin films~\cite{PhysRevB.69.024505,Baturina2004}. Vortex matching effect was studied in nanoperforated ultrathin TiN films~\cite{MIRONOV2010S808}. The emergence of ferromagnetism
has been discovered in TiN with an increase in nitrogen
vacancies~\cite{GUPTA2019221}.
Superconducting energy gap is an important parameter for superconducting materials, which can supply valuable information both for the physical understanding and the application of the materials.
This issue has been investigated through
scanning tunneling spectroscopy
(STS)~\cite{PhysRevLett.93.217005,PhysRevB.100.214505} and tunneling
junctions~\cite{9352504}. A strong spatial inhomogeneity with the gap value varying by a factor of 2 was found in TiN film~\cite{PhysRevB.100.214505}. The importance of
interaction between the SC films and substrates was noticed in recent years~\cite{Prischepa_2021,PRL147006,FeSe,Jiang_2024}.
Specific to the TiN thin film, it was found that type of substrates has a clear impact on the lattice orientations,
e.g., (111) TiN on $c$-cut
sapphire~\cite{doi:10.1021/acs.cgd.7b01278,PhysRevApplied.12.054001,doi:10.1021/acsphotonics.9b00617},
(110) TiN on $R$-plane sapphire~\cite{Torgovkin_2018}, and (100) TiN on
(100) magnesium oxide (MgO)~\cite{10.1063/1.4759019} and (100)
silicon~\cite{10.1063/1.107568,10.1063/1.3517252}. Meanwhile, disorder is another noteworthy factor, which can have a pronounced effect on the superconductivity in thin films~\cite{Anderson1983}.
At present, there is still a lack of comprehensive research on the superconducting behavior, especially in terms of superconducting gap, of TiN films with the variation of disorder level and crystal orientation.

In this work, we carried out the investigations on SC properties, especially the SC energy gap, of
TiN films grown on different substrates.  It
is found that the type of substrate has had a significant impact on both the crystal orientation and SC properties of TiN films.
Importantly, the TiN film grown on (100) silicon substrate displays a
significantly large SC gap which exceeds the prediction of weak-coupling BCS
theory, while that on (0001) sapphire and (100) MgO reveals a weak-coupling behavior. The analysis indicates that the disorder level and film orientation are not the most key factor in determining the superconductivity in the
present system. The influence of grain size seems to be more significant.

\section*{2. Experimental}
TiN films were grown using dc magnetron sputtering equipped with a
high vacuum pump.Three different substrates, (0001) sapphire (abbreviate as sapphire hereafter), high resistivity (100) silicon (abbreviate as Si hereafter), and (100) MgO were used.
The size of Ti target is 2 inches. The sputtering process was carried out in a
mixed atmosphere of Ar ($99.999\%$) and $N_2$ ($99.999\%$) with the flow rate of 15 sccm and 1.8 sccm respectively. The power and deposition pressure were fixed as 155 W and 0.16 Pa respectively for all the experiments.
For the case that sapphire was used, two different growth temperatures $T_{gr}$, 310 $^o$C and 500 $^o$C, were adopted. With
the Si and MgO substrates, TiN films were grown at 500 $^o$C. It is found that when the thickness of the film exceeds 40 nm, the surface morphology begins to exhibit island-like structures,
and its root mean square roughness will significantly increase. As for the samples with a much lower thickness, the XRD diffraction peaks are too weak, making it difficult to determine the crystal orientation.
Therefore, we choose the TiN films with a thickness of 35 nm for research in the present work. The detailed growth processes will be reported elsewhere.
The information for the samples studied in this work are summarized in Table 1.

The crystal structures of films were measured by X-Ray diffraction
(XRD, Bruker D8 Dicover). The surface morphology of TiN films was measured by atomic force
microscope (AFM, Bruker Dimension Icon). The thickness of TiN films was determined
by the X-ray reflectivity (XRR) measurements (Bruker, D8 Discover). The electrical
transport measurement were performed on a physical property
measurement system (PPMS, Quantum Design). The applied electric
current is 10 $\mu A$.
The superconducting gap of TiN films was studied by point-contact
spectroscopy. The point contact was achieved through
the nano-Au array on the films, see Fig. 5(a). The nano-Au array was synthesized on
film by using thermal evaporation through a nanoporous AAO mask
attaching to the surface of TiN film. The commercially purchased AAO films were
chosen to be the mask of the nano-Au array. The pore diameter of AAO film is 30
nm with the interpore distance of 60 nm and the thickness of 100 nm. The detailed processes for the fabrication of nano-Au array can be seen in our previous work~\cite{PRB174502}.

\begin{table}
\centering \caption{Summary of the information for the film growth and orientation.}
\medskip
\begin{tabular}{p{1.6cm}<{\centering}p{1.6cm}<{\centering}p{1.7cm}<{\centering}p{1.5cm}<{\centering}p{1.5cm}<{\centering}}
\hline
Name & Substrate &   $T_{gr}$ ($^o$C) & Orientation & $R_q$ (nm)\\
\hline
Sapp-310  & Sapphire  & 310   & (111) &  0.49  \\
Sapp-500  &  Sapphire  & 500  & (111)  &  0.70\\
Si-500  &  Si         & 500   & (100)   & 1.38\\
MgO-500  &  MgO         & 500  & (100)  &  0.14\\
\hline
\end{tabular}
\end{table}

\begin{figure}\centering
\includegraphics[width=9cm]{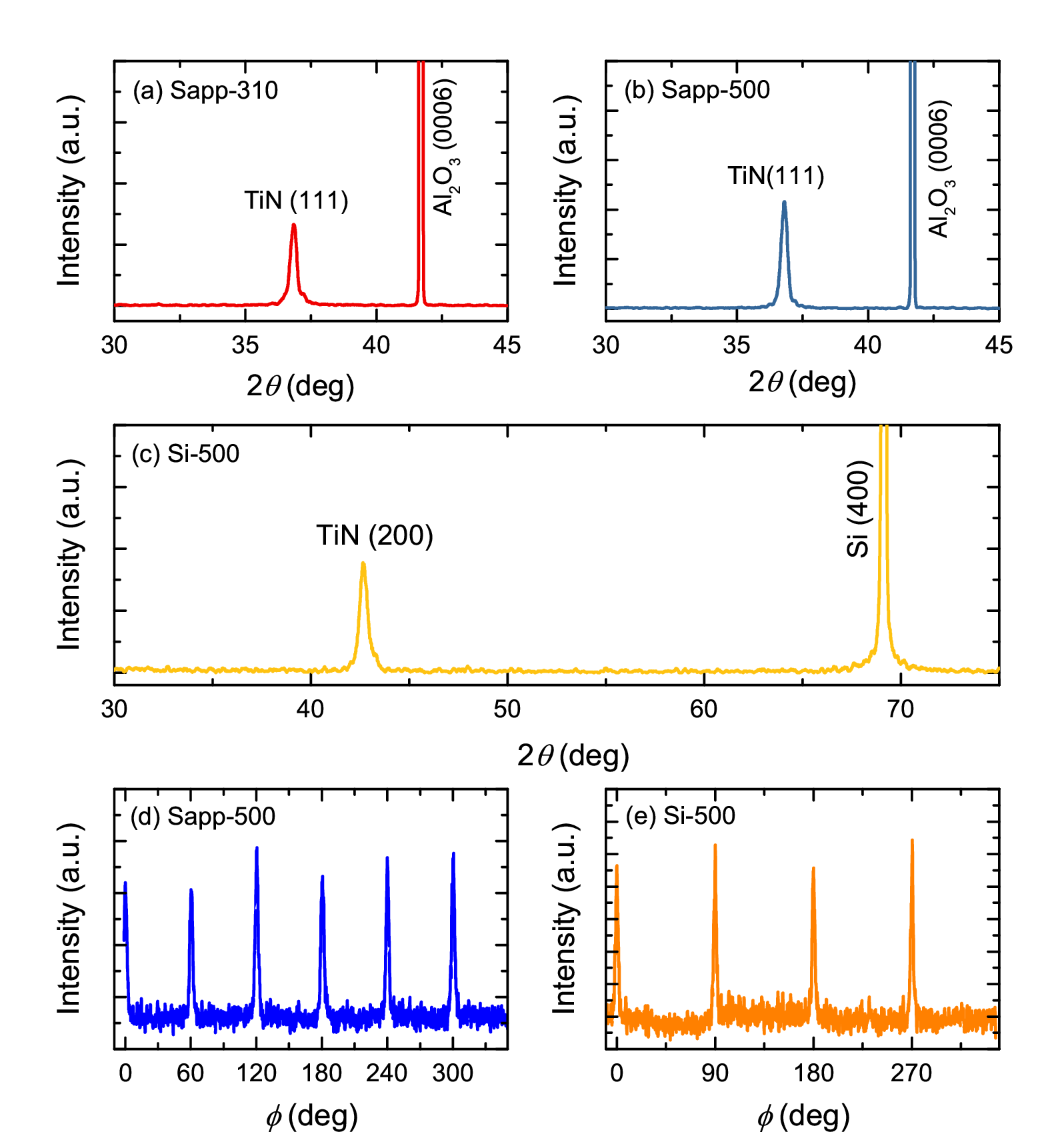}
\caption {XRD patterns of the three TiN films Sapp-310 (a), Sapp-500 (b), and Si-500 (c). (d) Azimuth $\phi$ scans of the off-axis (111) peak of Sapp-500 sample.
(e) Azimuth $\phi$ scans of the off-axis (200) peak of Si-500 sample.} \label{fig1}
\end{figure}

\begin{figure*}\centering
\includegraphics[width=14cm]{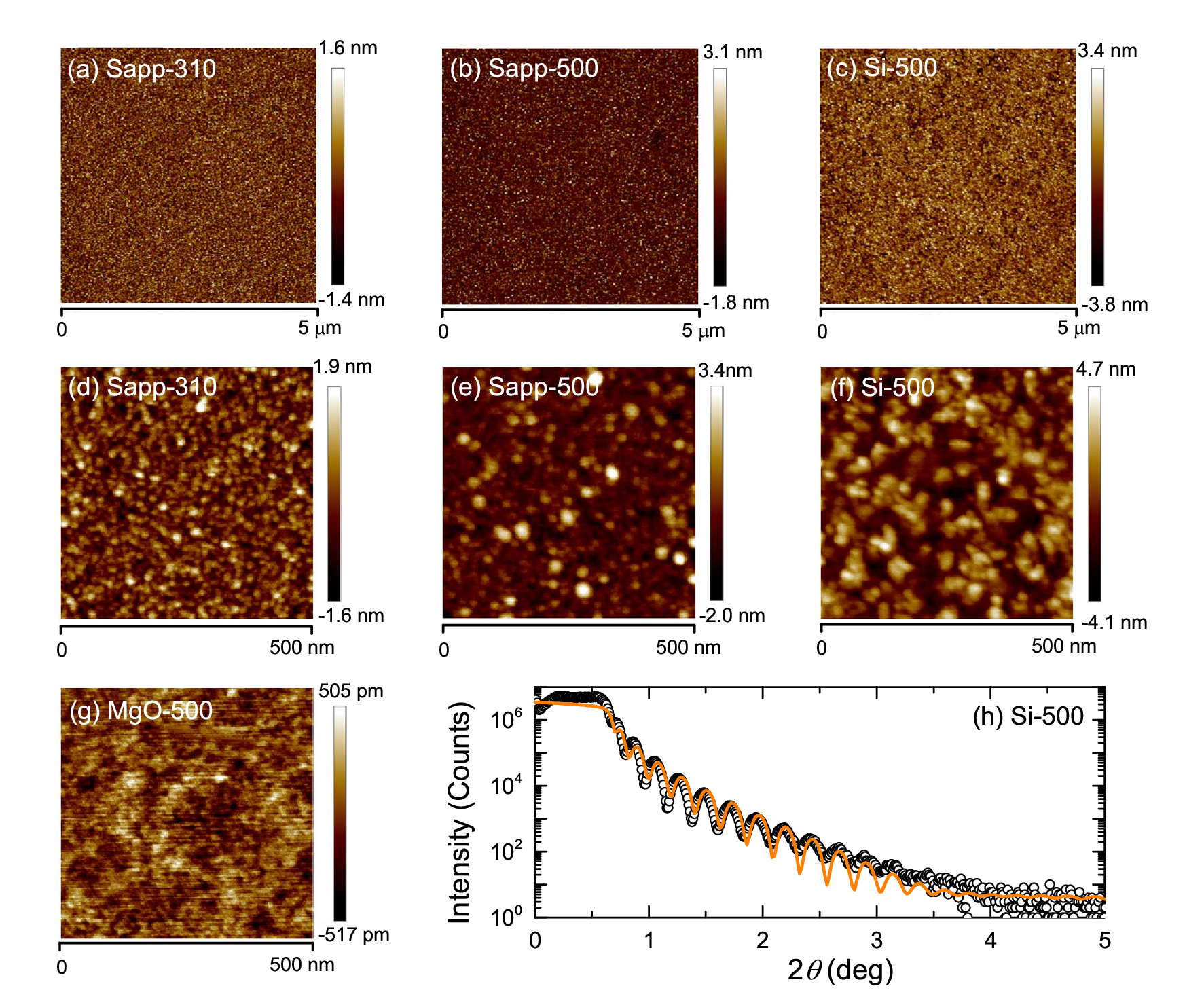}
\caption {(a-c) AFM images of the samples Sapp-310, Sapp-500, and Si-500, respectively. (d-f) The enlarged view of the AFM images of the three samples. (g) AFM image of the sample MgO-500.
(h) X-ray reflection curve for sample Si-500. The orange curve shows the fitting result.}
\label{fig1}
\end{figure*}

\section*{3. Results}
\subsection*{3.1 Characterization}
The crystal structure of the films were checked by XRD. Figs. 1(a, b)
show the XRD patterns of the two films grown on sapphire at different temperatures. The
peaks at around 36.8 $^\circ$ can be assigned to the (111) diffraction peaks of
TiN. No peaks from other orientations can be detected, indicating the oriented growth of the films.
For the film Si-500, only the (200) peak can be seen, see Fig. 1(c). Such substrate dependent film orientation is consistent with the previous report
of other groups~\cite{doi:10.1021/acs.cgd.7b01278,PhysRevApplied.12.054001,doi:10.1021/acsphotonics.9b00617,10.1063/1.107568,10.1063/1.3517252}.
To have a quantitative comparison, we
calculate the lattice constant of three films using the Bragg$'$s Law. The lattice parameters vary very little among these three sample, ranging from 4.22 ${\AA}$ to 4.24 ${\AA}$.

In order to check the in-plane texture of the films, the azimuthal $\phi$ scans of the off-axis (111) and (200) diffraction peaks of the sample Sapp-500 and Si-500 were performed. As shown in Fig. 1(d),
six peaks spaced by 60$^\circ$ are found in the sample Sapp-500, revealing a good six-fold in-plane symmetry of the (111) crystal plane. As for the sample Si-500, as shown in Fig. 1(e), the data reveals a clear
four-fold symmetry, being consistent with the feature of the (200) crystal plane. These results demonstrate the good in-plane orientation of the TiN films.

AFM images of samples Sapp-310, Sapp-500, and Si-500 are shown in Figs. 2(a-c). Within the scan range of 5 $\mu$m, the flat surface with a uniform distribution of the crystal grains can be observed.
By enlarging the picture to a range of 500 nm, the details of the crystal grains can be seen more clearly, see Figs. 2(d-f).
By comparing Sapp-310 and Sapp-500, it can be seen that the grain size becomes larger with the increase of $T_{gr}$. Meanwhile, under the same $T_{gr}$, TiN film grown on Si substrates shows larger grains.
Due to a better lattice matching between TiN and MgO, the TiN film on MgO reveals a much flatter surface with a small root-mean-square roughness $R_q$ of 0.14 nm. The values of $R_q$ of other samples
are summarized in Table 1.
The thickness of the films was obtained by fitting the XRR data using a
simple model with a four-layered structure (silicon substrate, SiN layer, TiN
film, and TiO layer). In Fig. 2(h), we show the XRR data (black circles) and the fitting result (orange curve) of one typical sample Si-500. The
thickness of TiN film was determined to be 35 nm.

\begin{figure}\centering
\includegraphics[width=7.5cm]{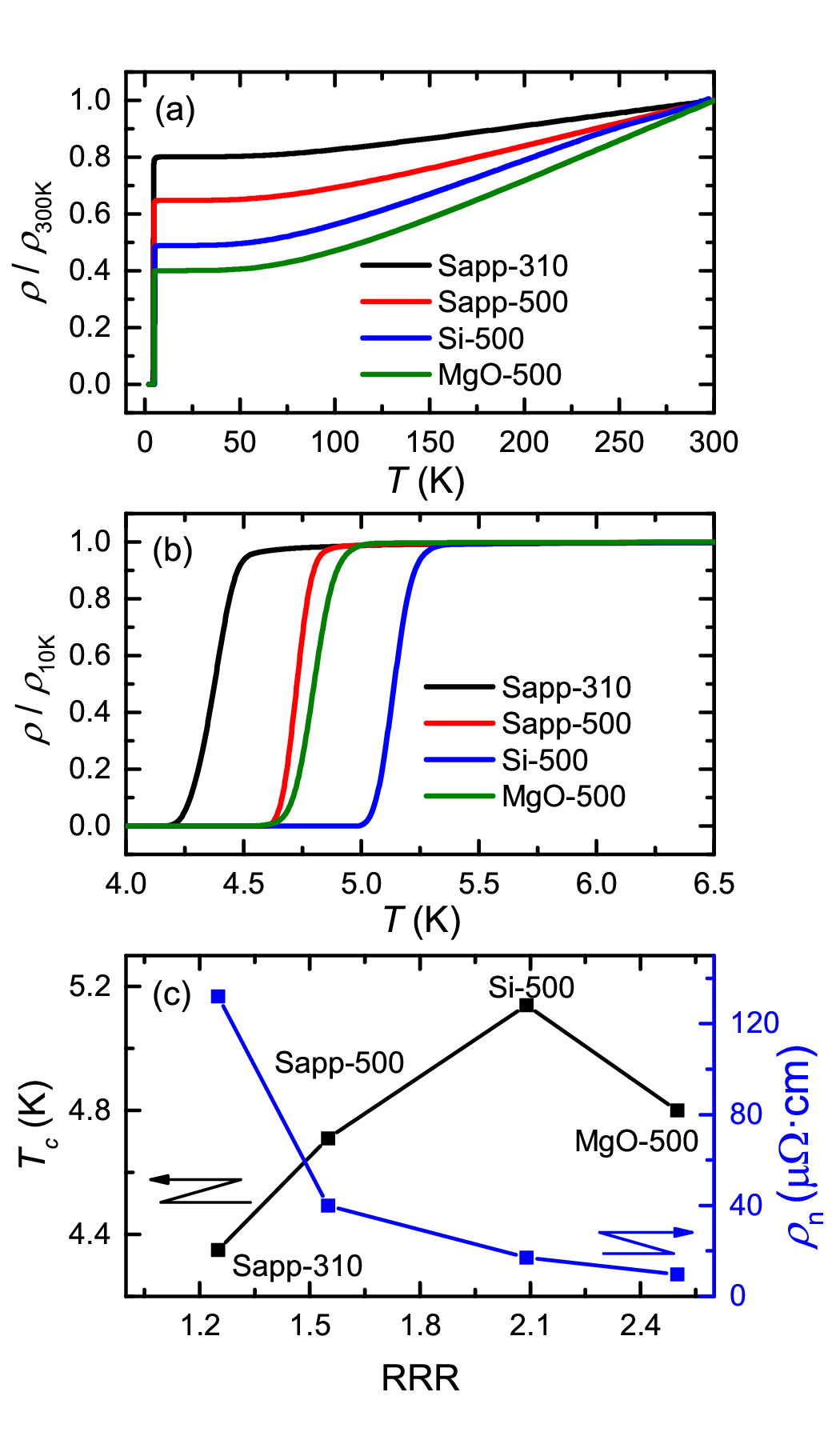}
\caption {(a) Normalized resistivity of the four samples as a function of temperature under
zero magnetic field. (b) An enlarged view of the normalized resistivity data in the low temperature. (c) SC critical temperature $T_c$ (left) and
normal-state resistivity $\rho_n$ (right) as a function of the residual resistivity ratio (RRR).} \label{fig1}
\end{figure}

\subsection*{3.2 Superconducting and normal-state properties}
Normalized resistivity of the four samples is shown in Fig. 3(a). All the samples reveal the SC transition in the low temperature region. The residual values of $\rho/\rho_{300K}$
show a systematic evolution with the growth temperature $T_{gr}$ and the type of substrate. In general, the residual resistivity ratio (RRR = $\rho_{300K}/\rho_{10 K}$) reflects the level of disorder in the film,
which contributes a temperature-independent term to the electrical transport. Thus the data indicates that a higher value of $T_{gr}$ can reduce the disorder level. Moreover, at the same growth temperature of
500 $^o$C, samples on Si and MgO substrate shows a even low disorder level as represented by the larger value of RRR. We note that the grain size and the inter-grain coupling via the grain boundaries
have a dominant influence on the magnitude of RRR~\cite{Ponta_2011}.
As shown in Fig. 2, TiN samples on the Si substrate display a clearly larger grain size as compared with that on sapphire. On one hand, a larger grain size will lead to the reduction of Coulomb energy.
Additionally, from a material perspective, larger grains mean fewer grain boundaries. This explains why the TiN film on Si exhibits a larger RRR.
Of course, in the case that the same substrate (i.e. sapphire) is used, higher growth temperatures result in larger grains that can also exhibit a larger RRR.
As for the sample on MgO, the very flat film surface and the small (almost invisible) grain boundaries result in a low level of disorder.

The SC transition can be seen more clearly in the enlarged view in Fig. 3(b).
The SC critical temperature $T_c$ displays a similar tendency to RRR with the variation of $T_{gr}$ and substrate except for that on MgO.
In the sample Si-500, $T_c$ can be as high as 5.14 K determined using the standard of 50\%$\rho_n$.
Here $\rho_n$ is the normal-state resistivity at the temperature just above the SC transition. In Fig. 3(c), we summarize the values of $T_c$ and $\rho_n$ as a function of RRR.
It is evident that, except for MgO-500, $T_c$ is positively correlated with RRR, while $\rho_n$ reveals a reverse evolution tendency. We note that this is a common feature in SC films~\cite{Liu_2022,WANG20231354223},
which actually reflects the suppression effect on superconductivity in low dimension induced by disorder.
The exceptional performance of sample MgO introduces some complexity, which will be analyzed in the Discussion section.

\begin{figure}\centering
\includegraphics[width=8.5cm]{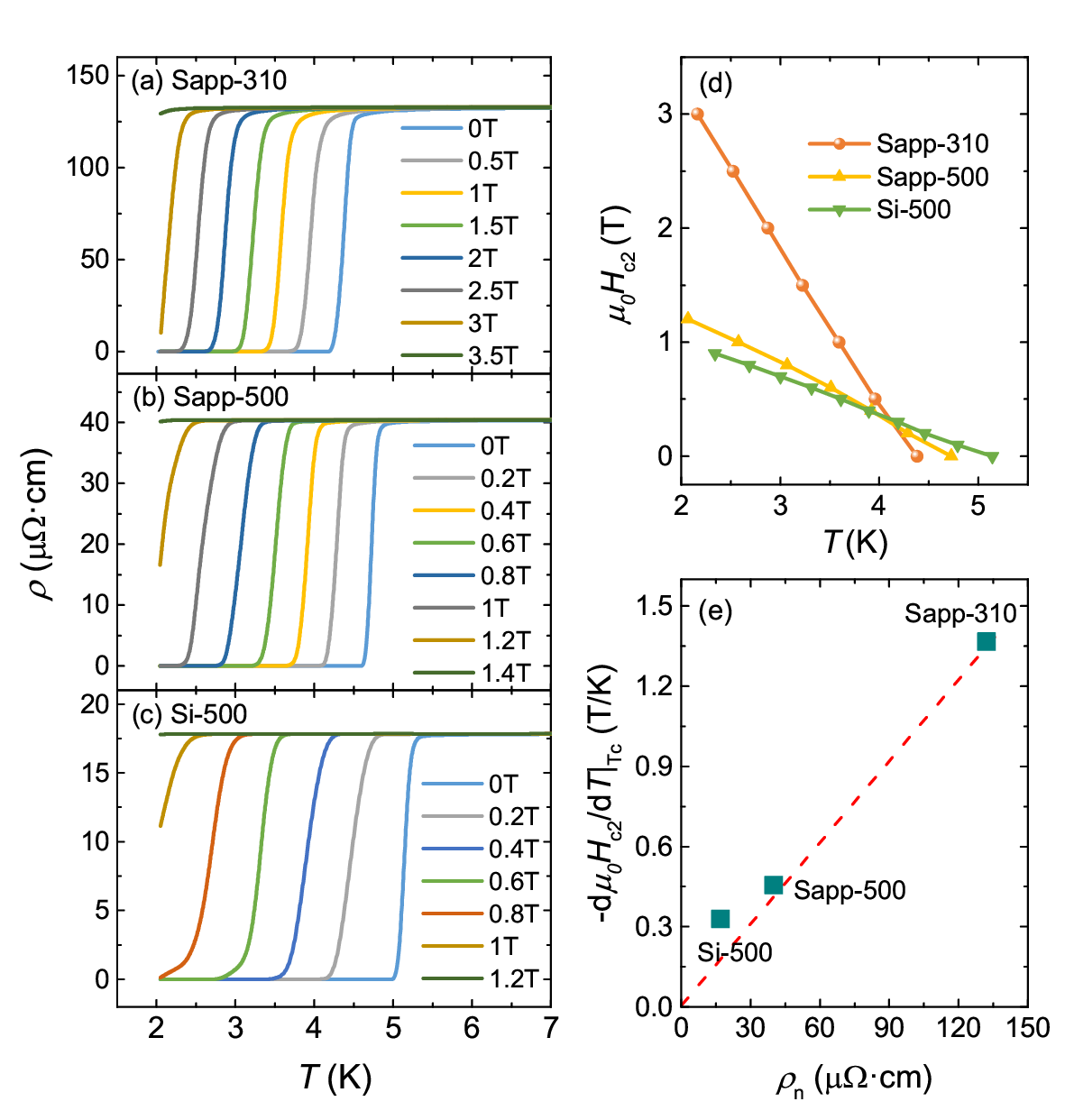}
\caption {(a-c) Temperature dependence of resistivity of the three samples under
the magnetic field. (d) Upper critical fields $\mu_0 H_{c2}$ as a function of temperature for the three samples. (e) Upper critical field slope near $T_c$ as a function of $\rho_n$.
The red dashed line is a guide for eyes.} \label{fig1}
\end{figure}

\begin{figure*}\centering
\includegraphics[width=16cm]{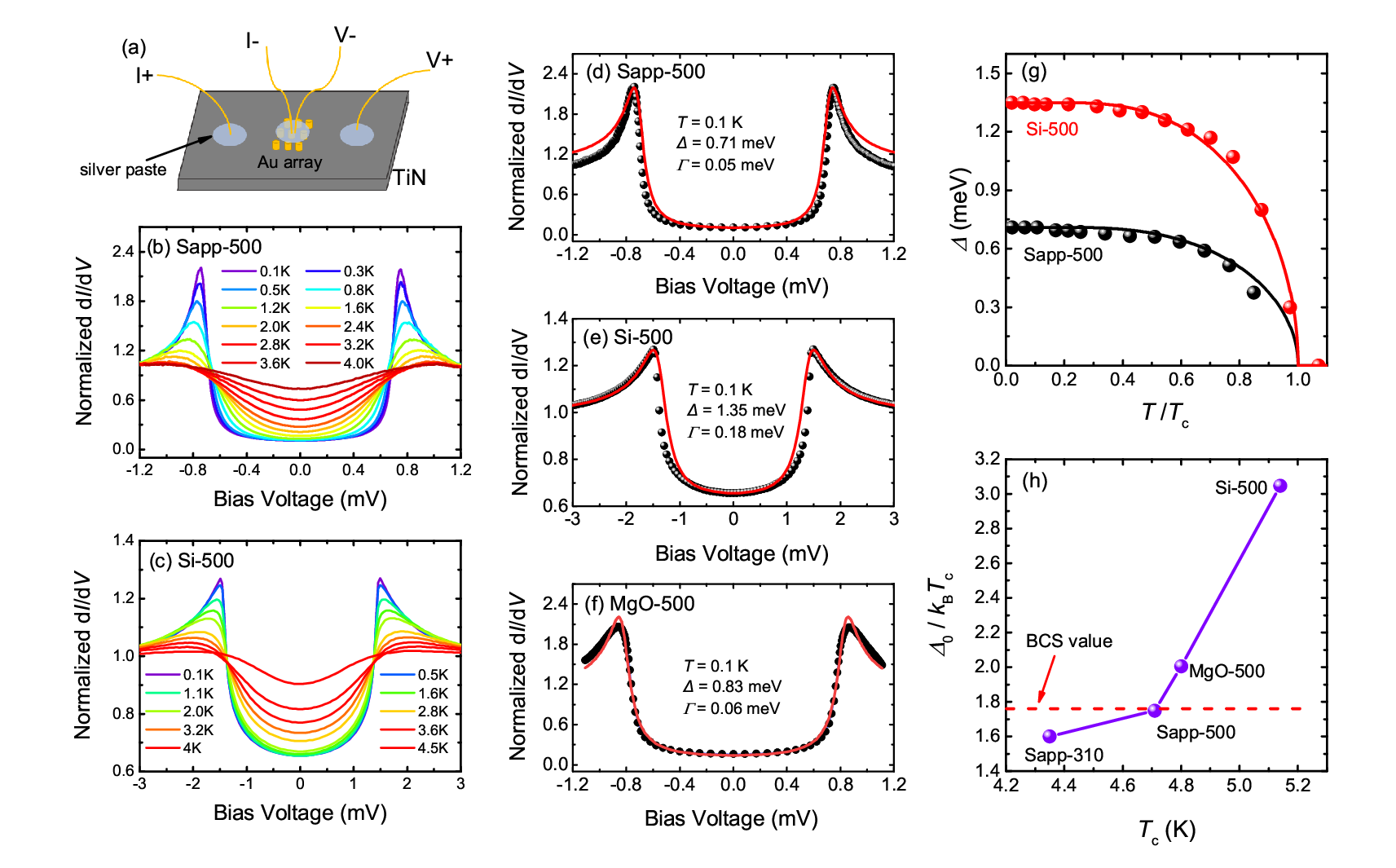}
\caption {(a) The schematic structure of the
measurement configuration for the point-contact tunneling spectrum.
(b, c) Temperature dependence of tunneling spectra of samples Sapp-500 and Si-500 respectively. (d-f) The spectrum data of the three samples at 0.1 K (black solid circles) with the best
fitting result (red lines) using the BTK model. (g) Temperature dependence of
the superconducting energy gap for the two samples. (h) The ratio $\Delta_0/k_BT_c$ as a function of $T_c$ for the four samples Sapp-310, Sapp-500, MgO-500, and Si-500.
The red dashed line shows the value predicted by the BCS theory.}
\label{fig1}
\end{figure*}

In Figs. 4(a-c), we show temperature-dependent
resistivity with various magnetic fields perpendicular
to the film surface for the three samples. As the field
increases, the SC transition shifts to lower temperatures.
It is notable that the suppression efficiency of magnetic field on superconductivity shows significant differences among these three samples. For the sample Sapp-310 with the
lowest $T_c$, superconductivity can survive under the field up to 3 T in the temperature range above 2 K. In contrast, superconductivity is suppressed completely by the field of 1.2 T for the sample
Si-500 with a higher $T_c$. Such a comparison can be seen more clearly in Fig. 4(d), where the temperature dependence of upper critical fields $\mu_0H_{c2}$ for all the three samples are shown.
In the temperature range above 2 K, $\mu_0H_{c2}$ display a linear decrease with the decrease of temperature. The slope of this tendency, $-d\mu_0H_{c2}/dT_c|_{T_c}$, is plotted in
Fig. 4(e) as a function of $\rho_n$. The roughly linear dependence of $-d\mu_0H_{c2}/dT_c|_{T_c}$ with $\rho_n$, as represented by the red dashed line, actually accords with the expectations
for dirty-limit superconductors~\cite{JaffePRB1989}.

According to the Werthamer-Helfand-Hohenberg relation\cite{Werthamer1966}
\begin{equation}
\mu_0H_{c2}(0)=-0.693\times d \mu_0H_{c2}(T)/dT|_{T_c}\times T_c, \label{eq:1}
\end{equation}
the zero-temperature upper critical field can be determined to be $\mu_0H_{c2}(0)$ = 4.2 T, 1.5 T, and 1.2 T for samples Sapp-310, Sapp-500, and Si-500, respectively.
The in-film coherence length was calculated to be 8.9 nm, 14.8 nm, and 16.6 nm for the three samples respectively.

\subsection*{3.3 Superconducting gap}
The point-contact tunneling spectra of TiN samples grown on different were measured. The point contact was realized by synthesizing
the nano-Au array on the films with the aid of nanoporous AAO mask. A schematic picture is shown in Fig. 5(a). The detailed experimental processes for the point-contact tunneling measurements
have been reported in our previous work~\cite{PRB174502}.

The obtained spectra for
samples Sapp-500 and Si-500 at a wide temperatures range down to 0.1 K are shown in Figs. 5(b) and (c).
The data are normalized to the spectrum at high temperatures where the superconductivity vanishes. Clear SC coherence peaks can be observed at low temperatures.
As the temperature increases, the broadening effect becomes apparent, and the differential conductivity curves gradually become flat.
The data are
analyzed using the Blonder-Tinkham-Klapwijk (BTK) theory~\cite{BTK}. The fitting results for the samples on three different substrates at the lowest temperature $T$ = 0.1 K are shown in Figs. 5(d-f).
For sample Si-500, in order to fit the data well, a term that remains constant with bias voltage is introduced. This may be due to the paralleling of other components in the point contacts.
The obtained SC gaps are $\Delta$ = 0.71 meV, 1.35 meV, and 0.83 meV for samples Sapp-500, Si-500, and MgO-500, respectively.
Meanwhile, the broadening parameters for the three samples are $\Gamma$ = 0.05 meV, 0.18 meV, and 0.06 meV, respectively.
The temperature dependence of gap value for Sapp-500 and Si-500 is summarized in Fig. 5(g), which is consistent with the BCS relation
(see the solid lines). We note that although the $T_c$ of sample Si-500 is slightly higher than that of Sapp-500, its SC gap value is significantly higher than the later.

To make a quantitative comparison,
we plot the ratio $\Delta_0/k_BT_c$ of four samples as a function of $T_c$ in Fig. 5(h). Here $\Delta_0$ is the extrapolated gap value in zero temperature, and $k_B$ is the Boltzmann constant.
It is found that the $\Delta_0/k_BT_c$ value of Sapp-310 is slightly lower that the weak-coupling BCS prediction of 1.76, while that of Sapp-500 is very close to this value.
Impressively, sample Si-500 exhibits a $\Delta_0/k_BT_c$ value as high as 3.05. The data of MgO-500 just follows the increasing tendency from that of Sapp-500 to Si-500.
Our result reveals an evolution from weak- to strong-coupling superconductivity in TiN films by changing the substrate from sapphire to Si.

\begin{table}
\centering \caption{Experimental results about the SC gap in TiN and NbN films.}
\medskip
\begin{tabular}{p{1.4cm}<{\centering}p{1.8cm}<{\centering}p{1.5cm}<{\centering}p{1.5cm}<{\centering}p{1.5cm}<{\centering}}
\hline
Sample & Substrate   & $\Delta_0$ (meV)    &  $\Delta_0/k_BT_c$ & Ref.  \\
\hline
TiN  &  Si$^a$  & 0.73            & 1.81  & \cite{PhysRevLett.93.217005}   \\
TiN  &   SiN/MgO$^b$  & 0.83            & 2.4  & \cite{9352504}   \\
TiN  &  Si           & 0.2-1.1            & 0.49-2.71  & \cite{PhysRevB.100.214505}   \\
NbN$^c$   & MgO          & 1.47-2.80        &  1.95-2.22   & \cite{PRB094509}     \\
TiN  &  Sapphire     & 0.60-0.71            & 1.60-1.75  & This work   \\
TiN   & Si           & 1.35           &3.05   & This work \\
TiN   & MgO           & 0.83           &2.01   & This work \\
\hline
\end{tabular}
\footnotesize{$^a$ The Si substrate was thermally oxidized.
$^b$ The specific type of substrate is not explicitly stated in the paper. $^c$ $T_c$ of NbN films covers a large range from 7.7 to 14.9 K.}
\end{table}

\section*{4. Discussion}
The SC gap of TiN films have also been studied by other groups. A gap value of 0.73 meV was reported in the magnetron sputtering sample grown on thermally oxidized Si substrates~\cite{PhysRevLett.93.217005}.
A slightly larger gap of 0.83 meV was detected in PLD-grown TiN film~\cite{9352504}. Notably, the experiments using STS measurements revealed a clear spatial inhomogeneity with the gap values ranging from
0.2 to 1.1 meV~\cite{PhysRevB.100.214505}. Meanwhile, another typical nitride superconducting material, NbN, can also be used as a valuable reference.
A complicated evolution of the gap ratio $\Delta_0/k_BT_c$ (ranging from 1.95 to 2.22) with $T_c$ and disorder level was uncovered in NbN films~\cite{PRB094509}. We summarize these result in
Table 2 to give a clear comparison with the present results.

\begin{figure}\centering
\includegraphics[width=8.5cm]{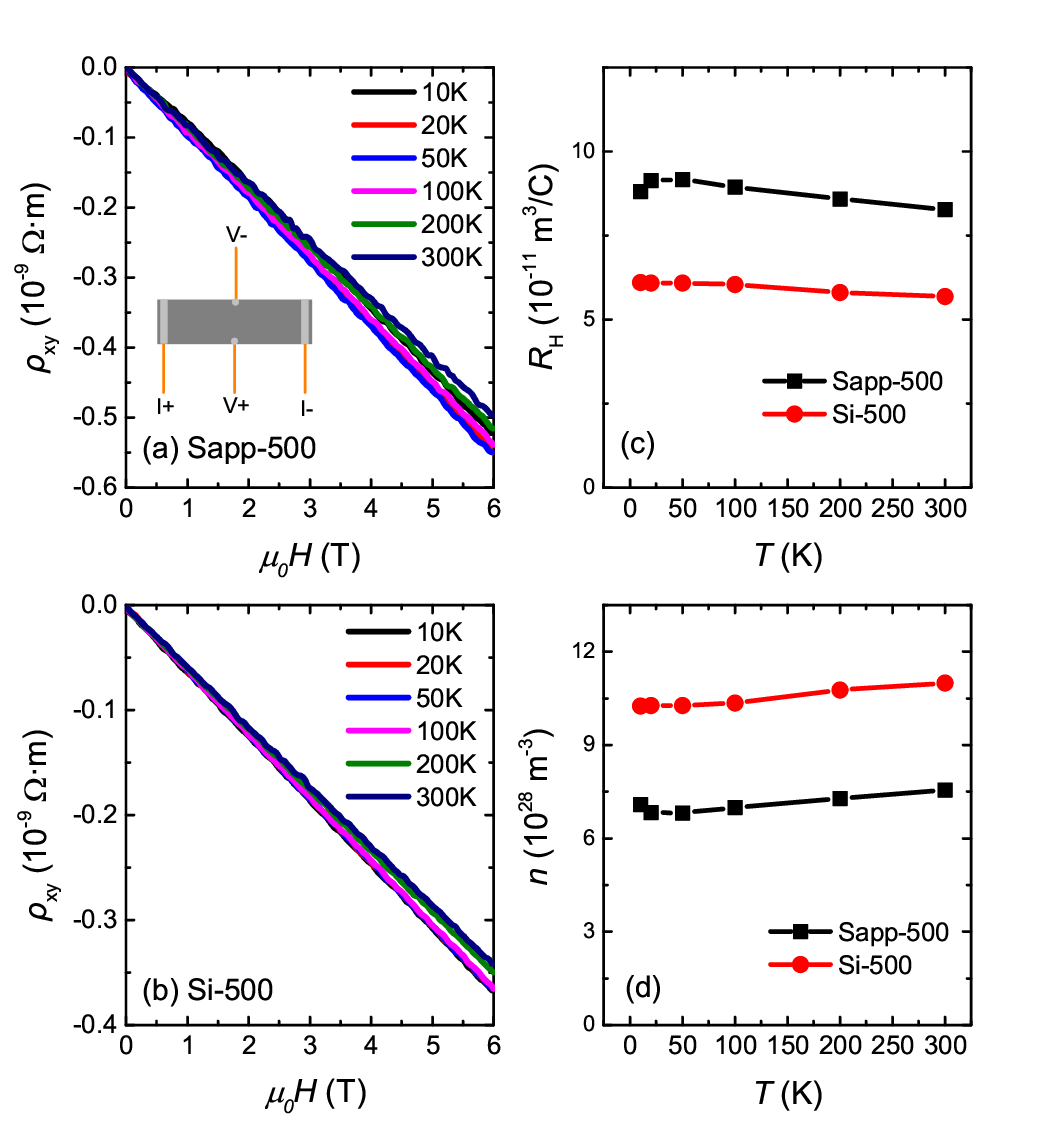}
\caption {Hall resistivity as a function of magnetic field of samples Sapp-500 (a) and Si-500 (b). The inset of (a) shows the schematic diagram of sample configuration for the Hall effect measurement.
Temperature dependence of Hall coefficient (c) and charge carrier density (d) of the two samples.} \label{fig1}
\end{figure}

Despite the significant divergence in experimental results regarding the SC gap of TiN thin films,
there is still a lack of further investigation into the physical origin behind it. Especially,
the underlying reasons for this change in coupling strength are worth exploring in depth. We first examined the lattice parameter changes caused by changes in substrate and growth temperature,
which directly affect the electronic structure of the material. As we have stated, the lattice parameter is very close among the three samples with the relative variation below 0.5\%.
Thus we argue that change in lattice structure is unlikely to be the main factor causing differences in the coupling strength.

\begin{table}
\centering \caption{Physical quantities calculated from the Hall and resistivity data.}
\medskip
\begin{tabular}{p{1.4cm}<{\centering}p{1.8cm}<{\centering}p{2.1cm}<{\centering}p{0.8cm}<{\centering}p{0.8cm}<{\centering}p{0.8cm}<{\centering}}
\hline
Sample & $n$      & $N(0)$               &  $\tau$ & $l$  & $k_Fl$\\
     & $e$/m$^{3}$ & states/m$^{3}$eV &     fs & nm   &  $-$  \\
\hline
Sapp-500  &   7.1$\times$10$^{28}$  & 1.70$\times$10$^{28}$  & 1.25  & 1.85 & 23.7   \\
Si-500   &   10.3$\times$10$^{28}$  & 1.92$\times$10$^{28}$  &1.95   & 3.28 & 47.6  \\
\hline
\end{tabular}
\end{table}

We know that disorder has a significant impact on superconductivity in low dimensional materials.
On the one hand, strong disorder weakens the electron screening effect in the system, thereby reducing the value of SC gap~\cite{Anderson1983}.
On the other hand, disorder can lead to electron localization, which thereby reduces the density of states at the Fermi level $N(0)$.
Actually, the data of $\rho_n$ and RRR in Fig. 2 have shown some information concerning the disorder levels.
To further check these two factors, we carried out Hall effect measurements on samples Sapp-500 and Si-500.
The sample was cut into rectangles with a dimension of 2 $\times$ 1 mm$^2$. The voltage terminals were fixed at both sides of the sample using silver paste to measure the Hall signal (see the inset of Fig. 6(a)).
As shown in Figs. 6(a) and (b), both samples reveal the linear behavior of the Hall resistivity $\rho_{xy}$ versus
magnetic field. Moreover, temperature has a rather weak influence on the Hall data.
The Hall coefficient $R_H$ and charge carrier density $n$ are calculated from the Hall resistivity data using the following equations
\begin{equation}
R_H = \rho_{xy}/\mu_0H, \label{eq:2}
\end{equation}
\begin{equation}
n = -1/eR_H, \label{eq:3}
\end{equation}
where $e$ is the charge of an electron. The results are displayed in Figs. 6(c) and (d) respectively.
One can see that the charge carrier density of sample Si-500 is about 50\% higher than that of Sapp-50, indicating a notable electron localization effect in TiN based on sapphire.
Assuming a spherical Fermi surface, the values of $N(0)$ can be derived using the relation
$N(0) = mk_F/\hbar^2\pi^2$, where $m$ is the mass of the electron, and $k_F = (3\pi^2n)^{1/3}$ is the Fermi wave vector. Employing the $n$ value at 10 K, N(0) is determined to be 1.70$\times$10$^{28}$ and 1.92$\times$10$^{28}$
states/m$^3$ eV for the two samples respectively.
Moreover, the relaxation time $\tau$ and mean free path $l$ of electron scattering have also been calculated combined with the resistivity data using the following equations
\begin{equation}
\tau = m/ne^2\rho, \label{eq:4}
\end{equation}
\begin{equation}
l = \hbar k_F\tau/m. \label{eq:5}
\end{equation}
The physical quantities calculated from the Hall and resistivity data are summarized in Table 3. The level of disorder is typically characterized by the Ioffe-Regel parameter $k_Fl$~\cite{PRB094509}.
It can be seen that compared to sample Si-500, sample Sapp-500 does indeed exhibit a substantially higher level of disorder.
Meanwhile, the density of states at the Fermi level of sample Si-500 is higher than that of Sapp-500.

Coulomb pseudo-potential $\mu^{\ast}$ and electron-phonon coupling parameter $\lambda$ are two important parameters in determining the superconductivity based on the McMillan theory~\cite{McMillan}.
It is notable that $\lambda$ is closely related to the density of states $N(0)$. Meanwhile, the presence of disorder can weaken the effective Coulomb screening effect, which
increases the value of $\mu^{\ast}$~\cite{Anderson1983,Zhang105008}. Based on the experimental results in the previous paragraph, the lower $T_c$ and weaker coupling strength of TiN on sapphire
substrate seem to come from the combined effect of enhanced electron localization and weakened Coulomb screening effect induced by disorder.

Although the increase in $N(0)$ and the decrease in disorder can qualitatively explain the strong-coupling feature exhibited in TiN films on Si substrates,
comparisons with other systems can provide more clues. According to the previous report~\cite{PRB094509}, the gap ratio $\Delta_0/k_BT_c$ of NbN system reveals a mild increasing with the disorder levels $k_Fl \geq$ 2.3.
At the same time, the values of $N(0)$ and $k_Fl$ display a greater increase~\cite{PRB214503} compared to the that between Sapp-500 and Si-500 studied in this work. In other words,
in the present TiN system, small changes in density of states and disorder are accompanied by a significant increase in energy gaps. Most importantly,
sample MgO-500 has the same orientation to that of Si-500 and shows an even low (high) level of $\rho_n$ (RRR), while it reveals a relatively
small value of $\Delta_0/k_BT_c$. Thus it seems that the film orientation and the disorder level are not the most key factors in determining the coupling strength.

Another noteworthy point is that, unlike STS measurements based on scanning tunneling microscopy~\cite{PhysRevLett.93.217005,PhysRevB.100.214505},
the point-contact tunneling method used in this work has a contact size of approximately 30 nm. Therefore, the information about the energy gap obtained in the present work is an average over a relatively large spatial scale.
In this case, the surface morphology of the thin films may have a non-negligible impact on the measurement results.
As shown in Fig. 2(f), sample Si-500 has a relatively large grain size ($\sim$40 nm). In this case, there is a high probability that the nano Au island used as point contacts will be located on an individual TiN grain.
In contrast, for samples Sapp-310 and Sapp-500, the TiN grains are clearly smaller than the Au islands and
each Au islands must cover an area containing a large number of grains and grain boundaries. Considering the fact that superconductivity at the grain boundaries is weakened, it seems rather reasonable
that the TiN films on sapphire reveal a weak-coupling feature. Furthermore, we noticed that despite the low level of disorder (low $\rho_n$ and high RRR),
$T_c$ of the sample on MgO was significantly lower than that on Si due to its smaller grain size.
These results indicate that grain size is an important influencing factor that cannot be ignored.

\section*{5. Conclusions}
Superconducting properties are investigated in TiN films grown on three different substrates. The SC critical temperature, normal-state resistivity and upper critical field all reveal
a systematically evolution with the disorder level. Strikingly, the system undergoes a transition from weak coupling ($\Delta_0/k_BT_c$ = 1.60 $\sim$ 1.75) to strong coupling ($\Delta_0/k_BT_c$ = 3.05),
accompanied by a change of the substrate changes from sapphire and MgO to Si.
Further analysis indicates that this change in superconducting coupling strength could not be fully explained by the disorder level or film orientations.
The surface morphology, especially the grain size, of the thin films may be an important factor influencing the superconductivity.

\section*{Data availability statement}
All data that support the findings of this study are included
within the article.

\section*{ACKNOWLEDGMENTS}
This work is supported by the Strategic Priority Research Program of the Chinese Academy of Sciences (No. XDB0670000),
the National Key Research and Development Program of China (Grant No. 2023YFB4404904), the Key-Area Research and Development Program of Guangdong Province, China (No. 2020B0303030002),
and the Autonomous Deployment Project of State Key Laboratory of Materials for Integrated Circuits (No. SKLJC-Z2024-B04).
The authors would like to thank all the staff at the
Superconducting Electronics Facility (SELF) for their
assistance.


\end{document}